\documentclass[12pt]{article}
\usepackage{psfrag,amsmath,euscript,amssymb,graphicx}

\begin{document}

\begin{titlepage}

\begin{flushright}
CLNS~04/1878\\
{\tt hep-ph/0405105}\\[0.2cm]
\today
\end{flushright}

\vspace{0.7cm}
\begin{center}
\Large\bf\boldmath
The Physics Case for an $e^+ e^-$ Super B-Factory
\unboldmath
\end{center}

\vspace{0.8cm}
\begin{center}
{\sc Matthias Neubert}\\
\vspace{0.7cm}
{\sl Institute for High-Energy Phenomenology\\
Newman Laboratory for Elementary-Particle Physics, Cornell University\\
Ithaca, NY 14853, U.S.A.}
\end{center}

\vspace{0.5cm}
\begin{abstract}
\vspace{0.2cm}\noindent
I share some personal reflections about the physics potential and
the physics case that can be made for an $e^+ e^-$ high-luminosity 
$B$-meson factory, as presented in my summary talk at the recent 
{\em Super $B$-Factory Workshop\/} jointly organized by the BaBar and 
Belle Collaborations (Honolulu, Hawaii, January 2004).
These brief remarks will appear as part of a forthcoming, comprehensive 
report on the physics potential of such a ``$10^{36}$~machine''.
\end{abstract}
\vfil

\end{titlepage}

\section{Introductory remarks -- Hopes and certainties}

The physics potential of an $e^+ e^-$ super $B$-factory must be
evaluated on the basis of a vision of the high-energy physics arena in
the 2010s. By that time, the BaBar and Belle experiments will
presumably have
been completed, and each will have collected data samples in excess of
500\,fb$^{-1}$. Hadronic $B$ factories will have logged several years
of data taking. There are excellent prospects that many parameters of
the unitarity triangle will have been determined with great precision
and in multiple ways. Likewise, many tests of the flavor sector and
searches for New Physics will have been performed using a variety of 
rare $B$ decays. A super $B$-factory operating at an $e^+ e^-$ 
collider with luminosity of order
${\cal L}\approx 10^{36}$\,cm$^{-2}$\,s$^{-1}$ would be the logical 
continuation of 
the $B$-factory program. If it is built, it will provide superb
measurements of Standard Model parameters and perform a broad set of
tests for New Physics. Such a facility could exhaust the potential of
many measurements in the quark flavor sector, which could not be done
otherwise. 

However, it cannot be ignored that a super $B$-factory
would come in the LHC era. By the time it could start operation, the
LHC will most likely (hopefully \dots) have discovered 
new particles, such as one or more Higgs bosons, SUSY partners of the
Standard Model particles, Kaluza--Klein partners of the Standard Model
particles, new fermions and gauge bosons of
a dynamical electroweak symmetry-breaking sector, or whatever else
will be revealed at the TeV scale. The crucial question is, therefore,
whether a super $B$-factory has anything to contribute to the physics
goals of our community in this era. More specifically, can it
complement in a meaningful way the measurements that will be performed
at the energy frontier? And while energy-frontier physics will most
likely attract most attention in the next decade or two, can a super
$B$-factory do fundamental measurements that could not be done
elsewhere (including earlier $B$-factories)? Would it be indispensable 
to our community's goal to comprehensively explore the physics at and 
beyond the TeV scale?

Fortunately, there exist indeed some big, open questions in
flavor physics, to which we would love to find some answers. Let me
mention three of them:

\paragraph{What is the dynamics of flavor?}
The gauge forces in the Standard Model do not distinguish between
fermions belonging to different generations. All charged leptons have 
the same electrical charge. All quarks carry the
same color charge. In almost all respects the fermions belonging to
different generations are equal -- but not quite, since
their masses are different. Today, we understand very little about the
underlying dynamics responsible for the phenomenon of generations.
Why do generations exist? Why are there three of them? Why are
the hierarchies of the fermion masses and mixing angles what they are?
Why are these hierarchies different for quarks and leptons? We have
good reasons to expect that the answers to these questions,
if they can be found in the foreseeable future, will open the doors to 
some great discoveries (new symmetries, forces, dimensions, \dots).

\paragraph{What is the origin of baryogenesis?}
The existential question about the origin of the matter--antimatter
asymmetry provides a link between particle physics and the evolution
of the Universe. The Standard Model satisfies the prerequisites for
baryogenesis as spelled out in the Sakharov criteria: baryon-number
violating processes are unsuppressed at high temperature;
CP-violating interactions are present due to complex couplings in
the quark (and presumably, the lepton) sector; non-equilibrium
processes can occur during phase transitions driven by the expansion
of the
Universe. However, quantitatively the observed matter abundance 
cannot be explained by the Standard Model (by many orders of
magnitude). Additional contributions, either due to new CP-violating
phases or new mechanisms of CP violation, are required. 

\paragraph{Are there connections between flavor physics and
TeV-scale physics?}
What can flavor physics tell us about the origin of electroweak
symmetry breaking? And, if the world is supersymmetric at some high
energy scale, what can flavor physics teach us about the mechanism of
SUSY breaking? Whereas progress on the first two ``flavor
questions'' is not guaranteed (though it would be most significant),
we can hardly lose on this third question! Virtually any extension of
the Standard Model that can solve the gauge hierarchy problem (i.e.,
the fact that the electroweak scale is so much lower than the GUT
scale) naturally contains a plethora of new flavor parameters. Some
prominent examples are:
\begin{itemize}
\item
SUSY: hundreds of flavor- and/or CP-violating couplings, even in the
MSSM and its next-to-minimal variants
\item
extra dimensions: flavor parameters of Kaluza--Klein states
\item
Technicolor: flavor couplings of Techni-fermions
\item
multi-Higgs models: CP-violating Higgs couplings
\item
Little Higgs models: flavor couplings of new gauge bosons ($W'$, $Z'$)
and fermions ($t'$)
\end{itemize}
If New Physics exists at or below the TeV 
scale, its effects should show up (at some level of precision) in
flavor physics. Flavor- and/or CP-violating interactions can only be
studied using precision measurements at highest luminosity. Such
studies would profit from the fact that the relevant mass scales will
(hopefully) be known from the LHC.

To drive this last point home, let me recall some lessons from the 
past. Top quarks have been discovered through direct production at the
Tevatron. In that way, their mass, spin, and color charge have been 
determined. Accurate predictions for the mass were available before,
based on electroweak precision measurements at the $Z$ pole, but also
based on studies of $B$ mesons. The rates for $B$--$\bar B$ mixing, as
well as for rare flavor-changing neutral current (FCNC) processes such 
as $B\to X_s\gamma$, are very sensitive to the value of the top-quark 
mass. More importantly, everything else we know
about the top quark, such as its generation-changing couplings
$|V_{ts}|\approx 0.040$ and $|V_{td}|\approx 0.008$, as well as its
CP-violating interactions ($\mbox{arg}(V_{td})\approx -24^\circ$ with 
the standard choice of phase conventions), 
has come from studies of kaon and $B$ physics. 
Next, recall the example of neutrino oscillations. The existence of
neutrinos has been known for a long time, but it was the discovery of 
their flavor-changing interactions (neutrino oscillations) that has
revolutionized our thinking about the lepton sector. We have learned
that the hierarchy of the leptonic mixing matrix is very different
from that in the quark sector, and we have discovered that
leptogenesis and CP violation in the lepton sector may provide an
alternative mechanism for baryogenesis.\\

In summary, exploring flavor aspects of the New Physics, whatever it
may be, is not an exercise meant to fill the Particle Data
Book. Rather, it is of crucial relevance to answer some fundamental,
deep questions about Nature. Some questions for which we have a 
realistic chance of finding an answer with the help of a 
super $B$-factory are:
\begin{itemize}
\item
Do non-standard CP phases exist? If so, this may provide new clues
about baryogenesis.
\item
Is the electroweak symmetry-breaking sector flavor blind (minimal
flavor violation)?
\item
Is the SUSY-breaking sector flavor blind?
\item
Do right-handed currents exist? This may provide clues about new gauge
interactions and symmetries (left--right symmetry) at very high energy.
\end{itemize}
I will argue below that the interpretation of New Physics signals at a 
super $B$-factory can be tricky. But since it is our hope to answer
some very profound questions, we must try as hard as we can. 

The super $B$-factory workshops conducted in 2003 at SLAC and KEK have 
shown that a very strong physics case can be made for such a machine.
During these workshops it has become evident (to me) that a
strength of a super $B$-factory is precisely that its success will
not depend on a single measurement -- sometimes called a ``killer
application''. Several first-rate discoveries are
possible and often likely. It is the breadth of possibilities and the
reach of a super $B$-factory that make a compelling physics case. 
As with electroweak
precision measurements, we can be sure that New Physics effects must
show up at some level of precision in flavor physics. The question
remains, at which level? In the ``worst-case scenario'' that we should 
not see any
large signals in $B$ physics, a super $B$-factory would play a similar
role as LEP played for our strive toward the understanding of 
electroweak symmetry breaking. It would then impose severe
constraints on model building for the post-LHC era.

\section{CKM measurements -- Sides and angles}

At a super $B$-factory, the goal with regard to CKM measurements is
simply stated: achieve what is theoretically possible! Many
smart theoretical schemes have been invented during the past two 
decades for making ``clean'' measurements of CKM parameters. 
We can safely assume steady theoretical advances in our field (the past
track record is impressive). This will lead to ever more clever
amplitude methods, progress in heavy-quark expansions and effective 
field theories, and perhaps breakthroughs in lattice QCD.
Unfortunately, all too often these theoretical proposals are limited by 
experimental realities. With a super $B$-factory, it would finally
become possible to realize the full potential of these methods. One of
the great assets of such a facility, which is particularly valuable in 
the context of precision CKM physics, is the availability of huge
samples of super-clean events, for which the decay of the ``other $B$ 
meson'' produced in $e^+ e^-\to b\bar b$ at the $\Upsilon(4S)$ is tagged 
and fully reconstructed. Full reconstruction costs a factor 1000 or so 
in efficiency, which demands super $B$-factory luminosities. Once 
statistics is no longer of concern, the reduction in systematic error
is a great benefit.

\subsubsection*{The sides \boldmath$|V_{ub}|$ and $|V_{td}|$\unboldmath}

A precision measurement of $|V_{ub}|$ with a theory error of  
5\% or less will require continued progress in theory. Determinations
from exclusive semileptonic $B$ decays need accurate predictions for
$B\to$\,\,light form factors from lattice QCD or effective field
theory. Determinations from inclusive $B$ decays need optimized cuts
and dedicated studies of power corrections in the heavy-quark
expansion. Recent advances using soft-collinear effective theory
appear promising, but there is still much work left to be done.
A super $B$-factory can provide vast, clean data samples of fully 
reconstructed decays, which would be an essential step toward
eliminating the background
from semileptonic decays with charm hadrons in the final state. It can
also yield high-precision data on the
$q^2$ dependence of form factors, and on the $B\to X_s\gamma$ photon
spectrum down to $E_\gamma\sim 1.8$\,GeV or lower. This would provide 
important constraints on theory parameters (e.g., shape functions). 

Another road toward measuring $|V_{ub}|$ is to study the leptonic 
decays $B\to\mu\nu$ or $B\to\tau\nu$, which would be accessible at a 
Super $B$-factory. The rates for these processes are proportional to
$f_B^2\,|V_{ub}|^2$. A lattice prediction for the $B$-meson decay 
constant can then be used to obtained $|V_{ub}|$. Alternatively, one 
can combine a measurement of the leptonic rate with that for the 
$B$--$\bar B$ mixing frequency to obtain the ratio 
$B_B^{-1/2}\,|V_{ub}/V_{td}|$, where the 
hadronic $B_B$ parameter would again have to be provided by lattice 
QCD. Such a determination would impose an interesting constraint on 
the parameters of the unitarity triangle.

A precision measurement of $|V_{td}|$ itself would require continued
progress in lattice QCD. Rare radiative decays (or rare kaon decays)
could also help to further improve our knowledge of this parameter.

\subsubsection*{The angles \boldmath$\beta=\phi_1$ and
$\gamma=\phi_3$\unboldmath} 

A super $B$-factory would allow us to exploit the full theory potential
of various methods for model-independent extractions of CP phases. We 
could finally do the measurements whose analyses require the least 
amount of theory input.
In the Standard Model, it's really all about $\gamma$ (the unique CP
phase in $B$ decays), in various combinations with $\beta$ (the CP
phase in $B$--$\bar B$ mixing). The importance of pursuing $\gamma$
measurements using different strategies (conventionally called 
measurements of $\alpha$ and $\gamma$) is that ``$\gamma$ 
measurements'' measure $\gamma$ in pure tree processes, whereas
``$\alpha$ measurements'' probe $\gamma$ in processes where penguins 
are present. Comparing the results obtained using these different 
methods probes for New Physics in penguin transitions, which are 
prominent examples of loop-induced FCNC processes in the Standard 
Model. The precision that can be
reached on $\beta$ and $\gamma$ using various techniques accessible at
a super $B$-factory is most impressive. A lot of marvelous physics 
can be done once such measurements will be at hand.

\section{Searching for New Physics -- Never stop exploring}

\subsubsection*{Probing New Physics with CKM measurements}

The path is clear. If different determinations of unitarity-triangle 
parameters would turn out to be inconsistent, then this would signal 
the presence of some New Physics. For instance, it is interesting 
to confront the ``standard analysis'' of the unitarity 
triangle, which is primarily sensitive to New Physics in $B$--$\bar B$ 
and $K$--$\bar K$ mixing, with mixing-independent 
constructions using charmless hadronic decays such as $B\to\pi K$, 
$B\to\pi\pi$, $B\to\pi\rho$, and others.
These studies, while not independent of theory, have already 
established CP violation in the bottom sector of the CKM matrix (the
fact that Im$(V_{ub})\ne 0$ with the standard choice of phase
conventions), while still leaving ample room for possible New Physics 
effects in $b\to s$ FCNC processes. 
(Some authors have argued that there are
already some tantalizing hints of New Physics in $b\to s$ transitions
sensitive to ``electroweak penguin''-type interactions.)

It is also interesting to confront different
determinations of $\beta$ with each other, such as the measurement of
$\sin2\beta$ from processes based on $b\to s\bar cc$ vs.\ $b\to s\bar
s s$ or $b\to s\bar q q$ (with $q=u,d$) quark-level transitions. 
One of the burning issues today is whether there is something real to the
``$\phi K_s$ anomaly'' seen by Belle, but not confirmed by BaBar. With
more precise data, many other decay modes can be added to obtain
interesting information and perform non-trivial tests of the Standard
Model.

Yet, let me stress that many more tests for New Physics can be done
outside the realm of CKM measurements. Several of those involve rare
hadronic $B$ decays. Others make use of inclusive decay processes. The
general strategy is to look for niches where the ``Standard Model
background'' is small or absent. One cannot overemphasize the importance
of such ``null (or close-to-null) measurements'', as they provide direct 
windows to physics beyond the Standard Model. In comparison, the
search for New Physics in CKM measurements always suffers from a 
large Standard Model background.

\subsubsection*{Probing New Physics in exclusive decays}

Rare (charmless) hadronic $B$ decays are usually characterized by the
presence of several competing decay mechanisms, often classified in
terms of flavor topologies (trees, penguins, electroweak penguins,
annihilation graphs, exchange graphs). These refer to the flow of
flavor lines in a graph but do {\em not\/} indicate the
possibility of multiple gluon exchanges. Therefore, reality is far
more complicated. Until a few years
ago, such nonleptonic decay processes were believed to be intractable
theoretically. This has changed recently, thanks to the advent of QCD
factorization theorems, perturbative QCD methods, and soft-collinear
effective theory, which complement previous approaches based on flavor
symmetries. Together, these approaches build the foundation of
a systematic heavy-quark expansion for exclusive $B$ decays,
much like heavy-quark effective theory provided the basis for such an
expansion in the (much simpler) case of exclusive $B\to D^{(*)}l\nu$
decays. 
(The dispute between QCD factorization and pQCD practitioners is also 
beginning to be resolved, since the issue of Sudakov logarithms in 
heavy-to-light transition amplitudes is now under good theoretical 
control.)

With ever improving theoretical control over exclusive $B$ decay
processes, several possibilities for tests for New Physics become
accessible. A partial list includes the measurement of $\sin 2\beta$
from the time-dependent CP asymmetry in $B\to\phi K_s$ decays, probing
electroweak penguins in rate measurements using $B\to\pi K_s$ decays,
and searching for New Physics by measuring CP asymmetries in 
$B\to K^*\gamma$ decays and the forward-backward asymmetry in 
$B\to K l^+ l^-$ decays. While there will always be an element of 
theory uncertainty left in these
analyses, in the cases above these uncertainties can be controlled
with rather good precision, so that large deviations from Standard
Model predictions would have to be interpreted as signs of New
Physics. (Indeed, some intriguing ``hints of anomalies'' are seen 
in present data.)

\subsubsection*{Probing New Physics in inclusive decays}

This is the more traditional approach, which profits from the
availability of reliable theoretical calculations. Several methods
have been discussed over the years, including precision measurements
of the $B\to X_s\gamma$ branching ratio and CP asymmetry, the
$B\to X_s\,l^+ l^-$ rate and forward-backward asymmetry, the inclusive 
$B\to X_s\nu\bar\nu$ decay rate, and some of the above with $X_s$ 
replaced with $X_d$. The mode $B\to X_s\nu\bar\nu$ is tough. This would
definitely be super $B$-factory territory.

\section{Interpreting New Physics -- The quest to measure non-standard 
flavor parameters}

The primary goal of a super $B$-factory would be to measure New
Physics parameters in the flavor sector. In
general, non-standard contributions to flavor-changing processes can be
parametrized in terms of the magnitudes and CP-violating phases of
the Wilson coefficients in a low-energy effective weak
Hamiltonian. The main obstacle is that, in general, there can be many 
such coefficients! Ideally, we would like to probe and measure
these couplings in a selective, surgical way, thereby measuring the 
fundamental coupling parameters of
new particles. Equally important is to study the {\em
patterns\/} of the New Physics, which may reveal important clues
about flavor dynamics at very high (beyond-LHC) energy scales.

\subsubsection*{CKM measurements}

A clean interpretation of New Physics signals in CKM measurements is
difficult (if at all possible) due to the large Standard Model
background. An important message is this: In the presence of New 
Physics, methods that are ``clean'' (i.e., that do not rely on theory 
input) in the Standard Model in general become
sensitive to hadronic uncertainties. This point is sometimes
overlooked. Consider, as an example, the
Gronau--London method for measuring $\gamma$ (or $\alpha$) from
$B\to\pi\pi$ decays. In the Standard Model, one needs five
measurements in order to extract the four unknown hadronic parameters
$|P/T|$, $|C/T|$, $\delta_{P/T}$, $\delta_{C/T}$ along with $\gamma$. 
With New Physics present, there are six additional amplitude 
parameters and not enough observables to fix them. But things are, 
in fact, worse than that, for the six new parameters are linear
combinations of New Physics parameters and a large number of hadronic
parameters -- the amplitudes and strong phases of the many $B\to\pi\pi$ 
matrix elements of the operators in the effective weak Hamiltonian. 
(It is a misconception to think that there is only one strong phase
each for the $\pi\pi$ final states with isospin $I=0$ or 2.)

The problem is, simply put, that CKM physics is hard. Consider how
difficult it has been (and still is) to determine the four parameters
of the CKM matrix, {\em for which there is no background}, since the
CKM matrix is the only source of flavor violation in the Standard
Model. With New Physics present, the Standard Model is a source of 
irreducible background for measurements in the flavor sector. In most
cases, the subtraction of this background introduces large hadronic 
uncertainties.

\subsubsection*{Non-CKM measurements}

In some cases, the Standard Model background can be strongly reduced
or even eliminated, so that one can directly probe certain types of 
New Physics operators. Examples are decay observables sensitive to
electroweak penguins, such as rate and CP asymmetry measurements in 
$B\to\phi K_s$ and $B\to\pi K_s$ decays. The 
idea is to look for certain patterns of ``isospin violation'', which 
in the Standard Model are highly suppressed, because they only arise 
at second order in electroweak interactions (``electroweak penguins''). 
This fact offers a window for seeing New Physics effects with little
Standard Model background. In many models, New Physics can fake the
signature of electroweak penguin operators {\em without\/} an
additional electroweak coupling involved (``trojan penguins''). This
provides sensitivity to sometimes very large energy scales (up to 
several TeV).
In other cases, such as $B\to VV$ modes or $B\to K^*\gamma$ decay, 
one can probe specific operators with non-standard chirality, thereby
eliminating the Standard Model background altogether. 

Searches for New Physics in inclusive decays are often simpler to
interpret, as they are afflicted by smaller theoretical
uncertainties in the relation between observables and Wilson
coefficient functions. Still, in general it can be difficult to
disentangle the contributions from (potentially many) new Wilson
coefficients, as only a limited number of observables can be 
measured experimentally. 

\subsubsection*{Importance of patterns of New Physics}

Let me close this discussion on an optimistic note. 
Even if it is hard to cleanly disentangle the contributions from
different New Physics operators, CKM measurements will play an
important role in helping to distinguish between different {\em
classes\/} of New Physics effects, such as New Physics in mixing vs.\ 
New Physics in
decay amplitudes, or New Physics in $b\to s$ vs.\ $b\to d$\, FCNC
transitions. CKM measurements might indicate the existence of new
CP-violating interactions or new flavor-changing interactions not
present in the Standard Model. Also, they will help to differentiate
between models with and without minimal flavor violation. 

Studies of exclusive hadronic decays can help to distinguish between
the ``flavor-blind'' transitions $b\to sg$ and $b\to s(\bar q q)_{\rm
singlet}$ and ``flavor-specific'' $b\to s(\bar q q)_{\rm non-singlet}$
decays. We will also be in a position to check for
the existence of right-handed currents and, more generally, probe for
operators with non-standard chirality.

\section{Conclusion}

Precisely because we don't know what to expect and what to look for,
it is the breadth of the physics program at a super $B$-factory that
will guarantee success. The discovery of new particles at the LHC would
help to interpret the possible findings of non-standard signals and 
guide further studies. Even finding no effects in some channels would
provide important clues.
Based on these consideration, it is my conviction that the physics
case for a super $B$-factory is compelling. Such a facility would be 
an obvious choice to pursue if any of the ``anomalies'' seen in the 
present $B$-factory data would ultimately turn out to be real effects 
of New Physics.

\subsubsection*{Disclaimer}

Above I have present some personal reflections about the physics 
potential and the physics case that can be made for a high-luminosity 
$e^+ e^-$ $B$-factory. My thinking about such a facility has evolved 
over a period of several years, starting with a workshop in June 2000 
in Glen Arbor, Lake Michigan that I helped organize. During this
process, I have 
profited from numerous discussions with colleagues. I have also been 
influenced significantly by the splendid performance of the SLAC and
KEK $B$-factories and of the BaBar and Belle experiments. Many things 
that were nearly unthinkable even a few years ago now appear within 
reach. (It is characteristic that the title of our 2000 Workshop
referred to a $10^{34}$ machine. In other words, the luminosity 
target has gone up by a factor 10 every two years!)

I have kept these introductory remarks brief. Much of the supporting
material will be presented 
in a forthcoming, comprehensive report on the physics potential of 
a ``$10^{36}$~machine''.

\subsubsection*{Acknowledgments}

I am grateful to Jon Rosner for valuable comments on the manuscript.
This work was supported by the National Science Foundation under Grant
PHY-0098631.

\end{document}